\documentclass[aps,prx,reprint,superscriptaddress,amsmath,amssymb,showpacs]{revtex4-2}

\usepackage{graphicx}
\usepackage{hyperref}
\usepackage{lipsum}

\par

\begin{document}

\title {Persistent short-range charge correlations revealed by ultrafast melting of electronic order in YBa$_2$Cu$_3$O$_{6+x}$}


\author{C. Seo}
    \affiliation{Linac Coherent Light Source, SLAC National Accelerator Laboratory, Menlo Park, CA 94025, USA}
\author{L. Shen}%
    \affiliation{Linac Coherent Light Source, SLAC National Accelerator Laboratory, Menlo Park, CA 94025, USA}
\author{A. N. Petsch}%
    \affiliation{Linac Coherent Light Source, SLAC National Accelerator Laboratory, Menlo Park, CA 94025, USA}
    \affiliation{Stanford Institute for Materials and Energy Sciences, SLAC National Accelerator Laboratory and Stanford University, Menlo Park, CA 94025, USA}
\author{S. Wandel}
    \affiliation{Linac Coherent Light Source, SLAC National Accelerator Laboratory, Menlo Park, CA 94025, USA}
\author{V. Esposito}
    \affiliation{Linac Coherent Light Source, SLAC National Accelerator Laboratory, Menlo Park, CA 94025, USA}
\author{J. D. Koralek}
    \affiliation{Linac Coherent Light Source, SLAC National Accelerator Laboratory, Menlo Park, CA 94025, USA}
\author{G. L. Dakovski}
    \affiliation{Linac Coherent Light Source, SLAC National Accelerator Laboratory, Menlo Park, CA 94025, USA}
\author{M.-F. Lin}
    \affiliation{Linac Coherent Light Source, SLAC National Accelerator Laboratory, Menlo Park, CA 94025, USA}
\author{S. P. Moeller}
    \affiliation{Linac Coherent Light Source, SLAC National Accelerator Laboratory, Menlo Park, CA 94025, USA}
\author{W. F. Schlotter}
    \affiliation{Linac Coherent Light Source, SLAC National Accelerator Laboratory, Menlo Park, CA 94025, USA}
\author{A. H. Reid}
    \affiliation{Linac Coherent Light Source, SLAC National Accelerator Laboratory, Menlo Park, CA 94025, USA}
\author{M. P. Minitti}
    \affiliation{Linac Coherent Light Source, SLAC National Accelerator Laboratory, Menlo Park, CA 94025, USA}
\author{R. Liang}
    \affiliation{Department of Physics and Astronomy, University of British Columbia, Vancouver, BC V6T 1Z1, Canada.}
    \affiliation{Quantum Matter Institute, University of British Columbia, Vancouver, BC V6T 1Z4, Canada}
\author{D. A. Bonn}
    \affiliation{Department of Physics and Astronomy, University of British Columbia, Vancouver, BC V6T 1Z1, Canada.}
    \affiliation{Quantum Matter Institute, University of British Columbia, Vancouver, BC V6T 1Z4, Canada}
\author{W. N. Hardy}
    \affiliation{Department of Physics and Astronomy, University of British Columbia, Vancouver, BC V6T 1Z1, Canada.}
    \affiliation{Quantum Matter Institute, University of British Columbia, Vancouver, BC V6T 1Z4, Canada}
\author{A. Damascelli}
    \affiliation{Department of Physics and Astronomy, University of British Columbia, Vancouver, BC V6T 1Z1, Canada.}
    \affiliation{Quantum Matter Institute, University of British Columbia, Vancouver, BC V6T 1Z4, Canada}
\author{C. Giannetti}
  \affiliation{Department of Mathematics and Physics, Università Cattolica del Sacro Cuore, Brescia, BS I-25121, Italy.}
\author{E. H. da Silva Neto}
    \affiliation{Department of Physics, Yale University, New Haven, CT 06520, USA}
    \affiliation{Energy Sciences Institute, Yale University, West Haven, CT 06516, USA}
    \affiliation{Department of Physics, University of California, Davis, CA 95616, USA}
\author{J. J. Turner}
    \affiliation{Linac Coherent Light Source, SLAC National Accelerator Laboratory, Menlo Park, CA 94025, USA}
    \affiliation{Stanford Institute for Materials and Energy Sciences, SLAC National Accelerator Laboratory and Stanford University, Menlo Park, CA 94025, USA}
\author{F. Boschini}
    \affiliation{Quantum Matter Institute, University of British Columbia, Vancouver, BC V6T 1Z4, Canada}
    \affiliation{Advanced Laser Light Source, Institut National de la Recherche Scientifique, Varennes, QC J3X 1P7, Canada}
\author{G. Coslovich}
    \email{gcoslovich@slac.stanford.edu}
    \affiliation{Linac Coherent Light Source, SLAC National Accelerator Laboratory, Menlo Park, CA 94025, USA}


%

\date{\today}
\begin{abstract}
Charge density waves (CDW) are ubiquitous in the complex phase diagram of cuprate superconductors and exhibit both short- and long-range correlations. Using time-resolved resonant X-ray scattering, we investigate the photo-induced dynamics of CDW in YBa$_2$Cu$_3$O$_{6.67}$. We discover an excitation threshold ($\Phi$$_\mathrm{C}$ $\approx$ 65 $\mu$J/cm$^2$) above which long-range CDW disappear, revealing a persistent CDW peak with short-range correlation length. Ultrafast photo-excitation promptly uncovers this residual short-range CDW correlations, appearing within $\approx$ 0.2 ps. Long-range CDW coherence recovers within $\approx$ 0.6 ps, while the peak intensity remains partially suppressed. We rationalize the dichotomic behavior in the fluence and temporal dependencies as the signature of two coexisting CDW peaks, arising from short- and long-range correlations, which we disentangle through their distinct response to photo-excitation. We provide evidence that the collapse of long-range correlations is driven by an electronic process, while short-range correlations are characterized by distinct timescales and stiffness against photo-excitation. This approach establishes ultrafast X-ray scattering as an effective tool for disentangling coexisting density waves and correlations in quantum materials.
\end{abstract}

\maketitle




The phase diagram of quantum materials emerges from electronic correlations and complex interaction between ordering phenomena, where subtle changes in external stimuli can dramatically alter the ground state \cite{fradkin2015,yin2022}. Among these phenomena, charge density waves (CDWs) have proven particularly significant, manifesting as periodic modulations of electron density, breaking the translational symmetry of the crystal lattice \cite{gruner1988}. In cuprate superconductors, the discovery of CDW order has revealed an intricate relationship with superconductivity that challenges the theoretical understanding \cite{keimer2015,castellani1995singular,caprara2017dynamical}. The coexistence and competition between these orders suggest that CDW formation may be intimately connected to the underlying mechanism of high-temperature superconductivity itself. However, the nature of this relationship, including how CDW respond to external perturbations and evolve across the phase diagram, remains a central unresolved question. Systematic experimental investigations across multiple tuning parameters have therefore become essential for unraveling these complex interactions.

Recent experiments have mapped the evolution of static and dynamic CDW correlations across various tuning parameters including doping, temperature, magnetic field, and pressure \cite{ghiringhelli2012,chang2012, gerber2015, comin2016, chang2016,jang2016, kim2018,choi2020, vinograd2024,porter2024understanding,arpaia2021charge,da2024dynamic}. Ultrafast spectroscopy has emerged as a powerful tool for investigating the temporal evolution of ordered states and their fluctuations \cite{giannetti2016ultrafast,zong2023emerging,boschini2024time}. However, while time-resolved optical spectroscopy provides insights into collective modes and electronic responses, it cannot directly access CDW at their characteristic wavevector \cite{torchinsky2013,hinton2013,dakovski2015}. Free-electron laser sources have overcome this limitation, enabling time-resolved X-ray scattering experiments that directly probe the CDW dynamics \cite{forst2014,bostedt2016linac,mitrano2019evidence,mitrano2019ultrafast,Wandel2022,jang2022,padma2025symmetry}.

\begin{figure}[!ht]
    \includegraphics[width=1\linewidth]{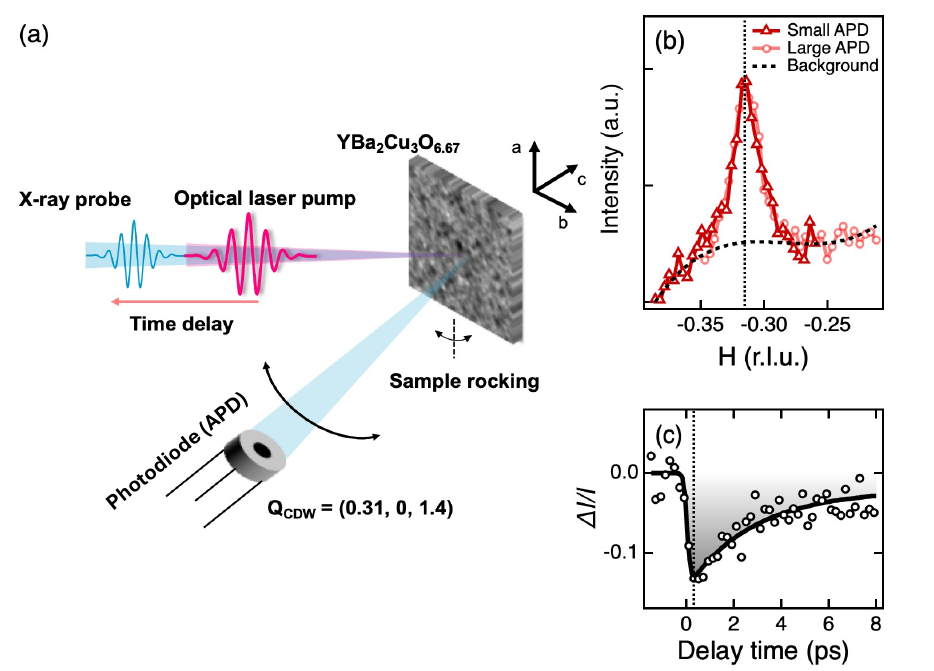}
    \caption{\label{fig:epsart} (a) Schematic diagram of the experiment. The 1.55 eV laser pump is utilized to perturb, while soft X-ray (931.5 eV) pulses probe the CDW of underdoped YBCO. (b) CDW peak profiles with small and large detector as a function of momentum before photo-excitation. (c) CDW peak dynamics as a function of delay time between pump and probe. The black vertical dotted lines in (b) and (c) mark the peak momentum and delays where the photoinduced signal is maximal, as considered in Figs. 2 and 3. }
\end{figure}

Time-resolved resonant soft X-ray studies have revealed complex interactions between superconductivity and charge order in YBa$_2$Cu$_3$O$_{6+x}$ (YBCO) \cite{forst2014,Wandel2022, jang2022}. When the superconducting state is transiently quenched by photo-excitation, the CDW correlation length shows a picosecond-scale, 90$\%$ increase, indicating the prompt annihilation of topological defects originally stabilized by superconductivity \cite{Wandel2022}. Additionally, precursor signatures of three-dimensional charge order were observed \cite{jang2022}. However, a comprehensive investigation of the ultrafast melting of CDW in YBCO remains unexplored.

Here we investigate the CDW dynamics in YBCO via time-resolved resonant soft X-ray scattering following optical excitation. Our experiment reveals distinct behaviors for long-range and short-range CDW correlations through systematic investigation of the time and fluence dependence. Long-range CDWs show a rapid melting process (within 0.2 ps), with a photo-induced signal reaching a saturation at relatively low fluence of $\Phi$$_\mathrm{sat}$ = 21.2 $\pm$ 2.3 $\mu$J/cm$^2$. Above a critical threshold fluence ($\Phi$$_\mathrm{C}$ $\approx$ 65 $\mu$J/cm$^2$), we observe a persistent residual broad peak with a correlation length of $\approx$ 27 \r{A}, comparable to the short-range correlations observed at equilibrium above the ordering temperature \cite{arpaia2019,arpaia2021charge,arpaia2023,Supplementary}. Recovery dynamics also shows signs of a dichotomic behavior for the CDW response. The peak width recovers within approximately 0.6 ps while the integrated intensity remains partially suppressed after strong pumping with a slower recovery of the amplitude. To explain the data, we elaborate a model based on two CDW components with completely different response to photo-excitation. We corroborate this two component interpretation by analyzing data collected with high momentum resolution, which allow to isolate the drop in amplitude of the long-range peak above the saturation fluence. 

The separation of timescales and fluence thresholds for short- and long-range CDW correlations in cuprates reveals a clear distinction in the nature of these two CDW components, with the long-range CDW collapse saturating at very low fluence, compatible with a purely electronic process, and recovering on ultrafast sub-picosecond timescale. Short-range CDW correlations show instead a markedly different stability under photoexcitation and distinct dynamics. The experimental findings reveal the unique ability of ultrafast X-ray techniques to disentangle complex and coexisting correlations in quantum materials through their dynamical response.

We performed time-resolved X-ray scattering experiments at the Linac Coherent Light Source (LCLS). The X-ray beam was tuned to the Cu L$_{3}$ absorption edge (931.5 eV) to resonantly probe the CDW signal in YBCO at \textbf{Q$_\mathrm{CDW}$} $\approx$ (0.31, 0, 1.4) \cite{schlotter2012soft}. The experimental geometry employed $\pi$-polarized X-rays, with the optical pump pulse polarization in the ab-plane. The experimental setup utilized a Ti:sapphire laser system delivering 50 fs optical pulses at 120 Hz. The pump beam was directed collinearly to the X-rays into the scattering chamber (Fig. 1a). During the experiment, we controlled the absorbed laser fluences in the range from 8 $\mu$J/cm$^{2}$ to 1.3 mJ/cm$^{2}$ \cite{Supplementary}. CDW peak mapping was achieved by rocking the sample through a 25$^{\circ}$ range while maintaining a fixed detector angle around 2$\theta$ = 160$^{\circ}$. Two photodiode detectors (APD) with different apertures were used, each separated by 10$^{\circ}$ in reciprocal space. The detector with smaller aperture had enhanced momentum resolution ($\Delta q$ = 0.002 $\mathring{A} ^{-1}$), at the expense of x-ray counts. Unless otherwise specified, the large acceptance detector was generally used in this work due to the higher signal to noise ratio. The CDW peak profiles measured by the two detectors are compared in Fig. 1(b), confirming that both detectors probe the same CDW signal.

Datasets were collected in two ways: first by tracking the temporal evolution of the CDW maximal intensity at \textbf{Q$_\mathrm{CDW}$} (Fig. 1c), and second, by recording the CDW diffraction profiles (theta scan) at a fixed time delay, typically the delay of maximal transient signal $\tau$$\approx$ 0.3 ps. The reference unpumped profiles were collected at negative time delays. All experiments were conducted at 65K, corresponding to the superconducting transition temperature  (T$_\mathrm{C}$ = 65K), where the CDW response is maximal. The CDW onset for this doping level is T$_\mathrm{CDW}$ = 160K \cite{ghiringhelli2012}. Notably, because no superconducting state is present at this temperature we expect no enhancement of the CDW peak from light-driven interaction with superconductivity \cite{Wandel2022,jang2022}, enabling us to study the intrinsic CDW melting dynamics in YBCO.

We first present the temporal evolution of the CDW response at various pump fluences. The time-resolved CDW dynamics show consistent temporal behavior across different fluences (Figures 2(a-c)). At fluence values of 25, 51, and 76 $\mu$J/cm$^2$, we observed the relative change in CDW peak intensity ($\Delta$I/I) as a function of pump-probe delay time. Using a single-exponential decay model convoluted with the laser-X-ray pulse cross-correlation, we observe an abrupt intensity decrease reaching maximum suppression at $\tau$ $\approx$ 0.3 ps, followed by a recovery with a characteristic timescale of approximately 2 ps. These timescales align with previous observations in YBCO and related charge-ordered materials, such as nickelates and tritellurides \cite{Lee2012,Zong2019,jang2020,Wandel2022,yusupov2010coherent}.

\begin{figure*}[!t]
    \includegraphics[width=0.95\linewidth]{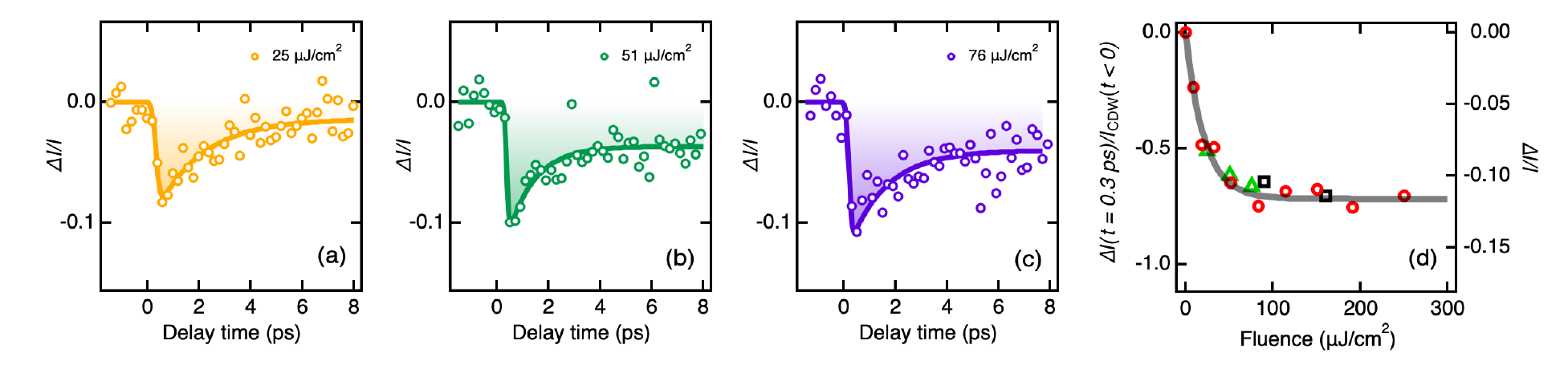}
    \caption{\label{fig:epsart} (a)-(c) CDW peak dynamics at three different fluence, 25, 51, and 76 $\mu$J/cm$^{2}$. The solid lines are the results of the fit of the data. (d) Fluence dependence of the normalized CDW peak decrease. Red circles represent data collected at a fixed time delay of 0.3 ps with varying pump fluence. Results are consistent with the peak values extracted from the time-domain scans in panels (a)-(c) (green triangles). Black squares represent similar data from Ref. \cite{Wandel2022}. The gray curve shows a saturation function fit to the data.}
\end{figure*}

Systematic fluence-dependence reveals a critical threshold behavior in the CDW response (Fig. 2(d)). The red circles represent direct measurements of CDW suppression at a fixed delay time ($\tau$ $\approx$ 0.3 ps) as a function of fluence. The green triangles show the maximum suppression values extracted from the time-domain data presented in Figures 2(a-c), while the black squares represent similar peak values obtained from previous measurements reported in Ref. \cite{Wandel2022}. The three datasets show a consistent saturation trend for the CDW response (Fig. 2(d)). 

The observed saturation in fluence indicates that the system is under strong perturbation, transitioning from a low fluence linear regime, where the CDW is suppressed proportionally to the incident energy, to a saturated state where the CDW state does not react to further photoexcitation. In charge-ordered systems \cite{Lee2012,Chuang2013real,beaud2014,esposito2017nonlinear} and superconductors \cite{kusar2008,stojchevska2011,coslovich2011evidence,giannetti2016ultrafast,Wandel2022}, such a saturation threshold in the fluence dependence is generally associated with the complete destruction of the order parameter within the probed volume. If we use a simple saturation model, of the form $\Delta$I=$1 - $$a(1-e^{-\Phi/\Phi_\mathrm{sat}})$, we obtain a saturation fluence, $\Phi$$_\mathrm{sat}$ = 21.2 $\pm$ 2.3 $\mu$J/cm$^2$. At this fluence the order parameter is melted at the surface of the photo-excited volume \cite{Supplementary}. We note that this saturation threshold is very low compared to other CDW systems \cite{tomeljak2009dynamics,padma2025symmetry}.

However, the evolution of the CDW momentum profile with fluence at $\tau$$\approx$ 0.3 ps reveals surprising behavior (Figures 3(a-e)). As the pump suppresses the CDW scattering intensity, the peak height gradually decreases, while the peak width remains mostly unchanged for low excitation fluences from the unpumped state. At a threshold fluence, $\Phi$$_\mathrm{C}$$\approx$ 65$\mu$J/cm$^2$, a dramatic change occurs, where the CDW peak exhibits a broader linewidth. This broad CDW peak persists through much higher fluences, where the temporal dynamics also becomes slower \cite{Supplementary}.

\begin{figure*}[hbt!]
    \includegraphics[width=1.0\linewidth]{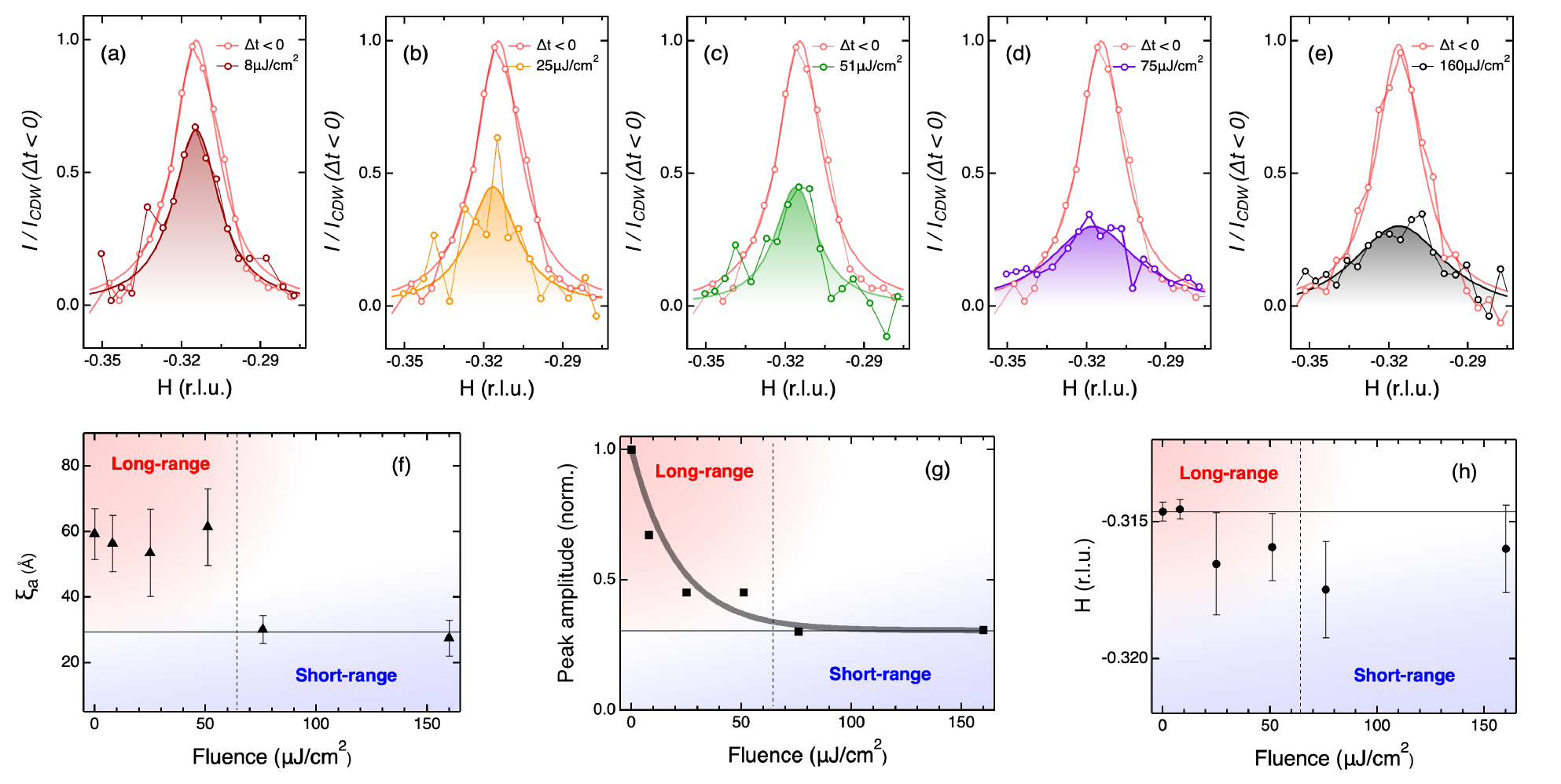}
    \caption{\label{fig:epsart} Fluence-dependent evolution of CDW characteristics. (a-e) CDW momentum profiles at $\tau$$\approx$ 0.3 ps for different pump fluences. All data are normalized to the unpumped CDW peak amplitude, following the subtraction of the fitted polynominal background as shown in Fig.1b. (f) Correlation length along the a-axis as a function of fluence. (g) Fluence dependence of the CDW peak amplitude. The thick dark gray  line is the same saturation function used in Fig.2(d). The red and blue shaded area indicate that the dominance of the narrow peak (long-range) and the broad peak (short-range). (h) Evolution of peak momentum vector as a function of fluence. Error bars represent one standard deviation derived from the fits in panels (a-e).}
\end{figure*}

The extracted correlation length along the a-axis ($\xi$$_a$) reflects this non-thermal behavior (Fig.3(f)). Rather than showing gradual changes with increasing fluence, $\xi$$_a$ maintains a relatively stable value around, 57 $\pm$ 3 \r{A}, throughout the low fluence regime, roughly matching the unpumped state. The system undergoes an abrupt transition at $\Phi$$_\mathrm{C}$, where $\xi$$_a$ sharply decreases to $\approx$27 \r{A}. This discontinuous change in correlation length contrasts sharply with the gradual evolution typically observed during thermal melting processes \cite{shen2026melting}. Notably, the reduced correlation length value is compatible with the short-range CDW correlations of $\approx$23 \r{A}, observed well above T$_\mathrm{CDW}$ via Resonant Inelastic X-ray Scattering (RIXS) in YBCO \cite{arpaia2019, arpaia2023,Supplementary}. 

This observation is particularly significant as it suggests that photo-excitation at this fluence level selectively disrupts long-range coherence while preserving short-range correlations, effectively separating these two charge ordering components via tr-REXS, without resorting to a challenging fit of two coexisting components in momentum space, or via RIXS \cite{arpaia2019,arpaia2023}. The persistence of a finite correlation length ($\approx$27 \r{A}) above the critical fluence reveals the robust nature of short-range charge correlations even when long-range order has been suppressed, highlighting a hierarchical structure in the charge ordering phenomenon and two distinct behaviors for the short- and long-range components of the charge order \cite{caprara2017dynamical}.

\begin{figure*}[hbt!]
    \includegraphics[width=0.9\linewidth]{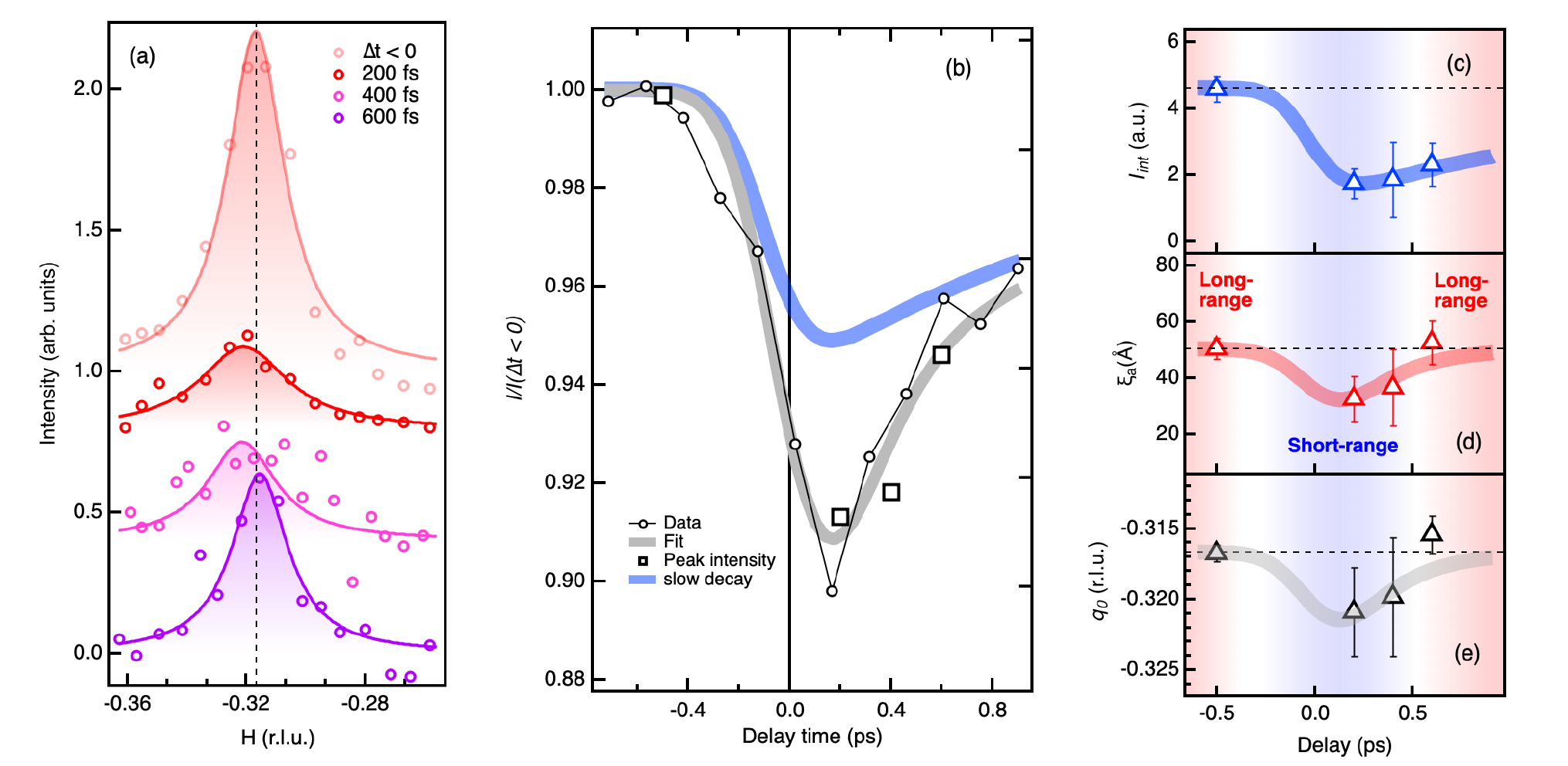}
    \caption{\label{fig:epsart} Temporal evolution of CDW dynamics following photo-excitation above the critical fluence threshold. (a) Momentum-resolved CDW scattering profiles at selected time delays relative to pump pulse arrival (t = 0). Color scheme: light pink (before time zero); red, pink, and purple curves represent profiles at 200 fs, 400 fs, and 600 fs post-excitation, respectively. (b) Time-resolved CDW peak intensity normalized to pre-excitation values. Black circles represent experimental data; thick gray line shows bi-exponential fit to the full temporal response. Thick blue line indicates the slow decay component isolated after subtracting the fast decay component from the fit. Black squares denote peak intensities extracted from momentum-resolved profiles in panel (a), demonstrating excellent agreement with time-domain trace at fixed $Q=Q_{CDW}$. (c) Integrated peak area, (d) correlation length, and (e) peak momentum position as functions of pump-probe delay. Thick colored lines represent temporal evolution predicted by fast (red, gray) and slow (blue) decay components derived from the bi-exponential analysis in panel (b), illustrating the distinct recovery dynamics of amplitude versus spatial coherence. Error bars represent one standard deviation derived from the fits in panel (a).}
\end{figure*}


The peak amplitude analysis as a function of fluence (Fig. 3(g)) shows behavior consistent with our intensity measurements presented in Figure 2(d), confirming the systematic suppression of CDW order with increasing fluence. The data clearly delineates two regimes: the dominance of a narrow peak (long-range) at low fluences, giving way to a broad peak (short-range) above $\Phi$$_\mathrm{C}$. We observe that the CDW peak almost maintains its characteristic wavevector across all measured fluences (Fig.3(h)), even as significant changes occur in both the peak width and amplitude, indicating the robustness of the ordering wavevector during this photo-induced transition.

To further understand the behavior above $\Phi$$_\mathrm{C}$, we measured the CDW peak dynamics at early delay times following photo-excitation, down to $\approx$ 100 fs \cite{Supplementary}. Figure 4(a) displays the evolution of CDW scattering profiles at various time delays after photoexcitation above $\Phi$$_\mathrm{C}$, progressing from the unpumped state through successive time points. To quantitatively analyze the temporal response with more time data points, we also collected the CDW peak intensity at $Q=Q_{CDW}$ as a function of delay time (Fig. 4(b)). The agreement between the time-domain trace and the peak amplitudes extracted from the momentum-resolved profiles in panel (a) (black squares) validates the internal consistency and robustness of the measurements. 

The temporal evolution exhibits distinctive bi-exponential behavior, as captured by the fitted curve (gray line). This analysis reveals two characteristic timescales: a rapid component with $\tau$ $\approx$ 0.6 ps and a slower relaxation extending beyond 3 ps. The decomposition of these components, with the slow decay contribution highlighted by the blue shaded line, provides essential insight into the steps needed for the CDW recovery dynamics. Importantly, we notice that the narrow peak associated with long-range CDW reappears around 0.6 ps delay, in synchronicity with the fast recovery dynamics.

The relationship between the two relaxation timescales and different parameters of the CDW peak profiles is further corroborated in Figs. 4(c-e). The integrated peak intensity exhibits an initial sharp suppression followed by a gradual ps-scale recovery (Fig. 4(c)), while both the correlation length and peak momentum position show remarkably rapid restoration to its pre-pump value within $\approx$ 0.6 ps (Fig. 4(d-e)). The guide-to-eye curves in panels (c) and (d) are derived from the fast and slow components of the decay functions identified in panel (b). This finding reveals a fundamental dichotomy: the fast relaxation dynamics is predominantly associated with the recovery of correlation length and peak position, whereas the slow component follows the amplitude recovery. 

To interpret this dichotomy we elaborate a model based on two CDW components representing short- and long-range order respectively, as schematically illustrated in Figure 5(a), with the long-range order amplitude dropping to zero around $\Phi$$_\mathrm{C}$ and a coexisting short-range order persisting up to much higher excitation fluence. The model, which is described in more detail in the supplementary material, qualitatively reproduces the trends seen in the experimental data as a function of fluence and time (Figs. 3(f) and 3(g))\cite{Supplementary}. The model reproduces the bi-exponential relaxation dynamics as well as the sudden discontinuities in the fluence dependence. Above $\Phi$$_\mathrm{C}$, the scattering properties, correlation length and peak position, abruptly collapse to the values of the underlying short-range order\cite{Supplementary}. 

Our two component model also captures the two key experimental observations from the temporal dependence: the rapid recovery of the correlation length and the persistent reduction in the integrated scattering intensity. The premise of the model, i.e., the presence of short- and long-range charge correlations, is in line with previously mentioned RIXS results in YBCO and other cuprates \cite{arpaia2019,arpaia2023,arpaia2021charge,chaix2017dispersive,Silva2018coupling,yu2020unusual,miao2017precursor,miao2019formation,li2020multiorbital,lee2021spectroscopic,boschini2021dynamic,huang2021quantum,lu2022identification,scott2023low}. New information is here conveyed by the extreme dichotomy of their response to photo-excitation, which allows discrimination of these two coexisting components thanks to the ultrafast probe.
In contrast, a single CDW picture would require assuming 1) a discontinuous evolution of the correlation length, with a significant broadening happening right at $\Phi$$_\mathrm{C}$ and 2) a sudden increase of stability to photo-excitation for the CDW as a function of correlation length.

To further corroborate this two-component interpretation, beyond the specific details of the model adopted, we considered the scattering data collected by the small detector in our experiment. Such detector, while having lower x-ray counts, can provide the higher momentum resolution to isolate the long-range (narrower) CDW contributions. This is depicted in the inset of Fig. 5(b), where the calculated 2D peak is shown in respect to the apertures of the two detectors in reciprocal space. The small detector here allows to increase the contrast to just the long-range contribution, thus isolating its response and clarifying whether a large broadening is happening above $\Phi$$_\mathrm{C}$. To compensate the loss of x-ray counts we integrate together 3 different fluences above $\Phi$$_\mathrm{C}$ and subtract the profiles without laser excitation.

The result is shown in Fig. 5(b), where the differential profile shows a clear drop of amplitude of the long-range order. The data can be fitted very well with a simple drop of amplitude while a model based on broadening does not reproduce the experimental data. This directly validates our two-component interpretation in the fluence regime across $\Phi$$_\mathrm{C}$, confirming that the long-range order is completely quenched around $\Phi$$_\mathrm{C}$ following its saturation. Interestingly, within this scenario we recognize that topological defects of long-range CDW would remain largely unperturbed by direct photo-excitation at these low fluence levels, in agreement with similar results in LESCO \cite{bluschke2024orbital}, thus allowing rapid, sub-ps scale, re-establishment of spatial order. 

Having discerned two separate components of the CDW, we analyze the stiffness of long-range CDWs against photo-excitation. We note that $\Phi_{sat}$, the melting fluence for the long-range CDW represents a small excitation energy, amounting to an energy density of just $\approx$ 0.8 meV/Cu, or equivalently 2.5 J/cm$^3$. Such energy density is smaller compared to typical melting energies in other CDW compounds\cite{tomeljak2009dynamics, Lee2012, padma2025symmetry}; for example, Peierls systems like TbTe$_3$ and K$_{0.3}$MoO$_3$, are characterized by a melting energy density of $\approx$ 3 meV/Tb and $\approx$ 4 meV/Mo respectively \cite{stojchevska2011, schmitt2008transient, tomeljak2009dynamics,huber2014coherent}. We observe instead that the CDW melting energy we obtain for the long-range CDW is similar to the energy density needed to disrupt the long-range superconducting order in YBCO \cite{stojchevska2011,boschini2018}. 

Quantitatively, if we were to consider the full lattice contribution to the specific heat and a transition from to 65K to $T_{CO}$, we would estimate a melting energy of $U_M$ = $\int_{65K}^{160K} C(T) \,dT\approx$ 158 J/cm$^3$. Instead the measured energy threshold is of the order just 1.6$\%$ of this value, around 2.5 J/cm$^3$. The small melting energy suggests that the melting of long-range order is an electronic process, and does not involve a significant phonon population, which would otherwise constitute a large heat bath with an out-sized component to the melting energy. For comparison, typical contribution from strongly coupled phonons can amount to around 20$\%$ of the full lattice contribution \cite{dal2012}.  This is consistent with the picture of charge order melting being governed by electronic energy scales rather than lattice instabilities, as evidenced by spectroscopic signatures of quantum critical charge fluctuations in optimally doped cuprates \cite{lee2021spectroscopic}.

\begin{figure}[t!]
    \includegraphics[width=1\linewidth]{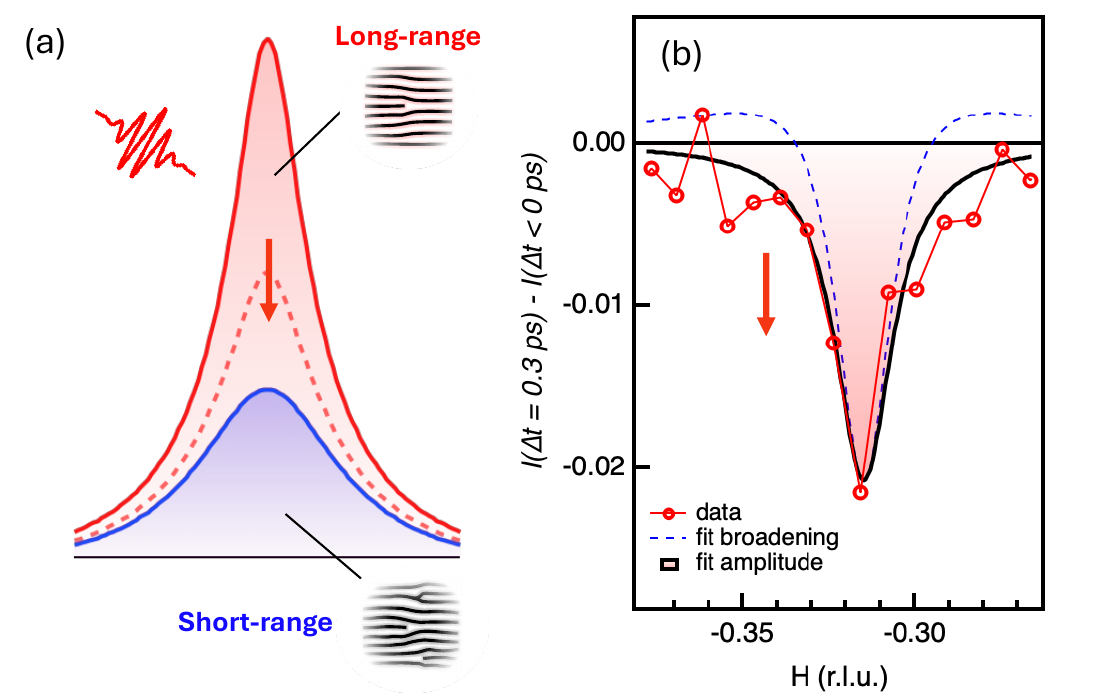}
    \caption{\label{fig:epsart} (a) Schematic evolution of long- (solid red line) and short-range (solid blue line) CDW contributions to the scattering peak using the two-peak model discussed in the main text and in the SI \cite{Supplementary}. The effect of photo-excitation is primarily to decrease the amplitude of the long-range order (red dashed line). (b) Differential change of the CDW scattering profile measured with enhanced momentum resolution. Red circles denote the experimental data, black line is the best fit with a differential model where the CDW amplitude is decreased following photo-excitation. The dashed blue line shows the case of a differential model where the CDW peak broadens instead. Full data and model in Ref.\cite{Supplementary}.} 
\end{figure}

Furthermore, if we estimate the heating effect at $\Phi$$_\mathrm{sat}$, the fluence at which long-range CDW order starts collapsing at the surface of the photo-excited volume, we obtain a global thermal increase $\Delta T_\mathrm{lat}\approx$ 2.6 K. Clearly, such small heating is insufficient to cause any collapse of CDW, which is rather a non-thermal process, with the electronic system being excited to much higher temperatures\cite{giannetti2016ultrafast}. The electronic temperature increase we estimate at $\Phi$$_\mathrm{sat}$ is of the order $\Delta T_\mathrm{el}\approx$ 100 K, very close to $T_\mathrm{CO}$ for this compound \cite{Wandel2022, ghiringhelli2012}, and thus compatible with a scenario where the collapse of long-range order is predominantly driven by the excitation of the electronic sub-system \cite{maklar2021nonequilibrium}. The fact that the collapse of long-range charge order is mostly driven by an electronic process is also supported by the observed relaxation dynamics. In fact, the rapid drop and recovery of peak scattering signal within 0.6 ps also indicates an electronic origin for the long-range order collapse, as these timescales are not compatible with the typical ps-scale anharmonic decay of strongly coupled phonons \cite{Lee2012,Zong2019}. Ultrafast probes sensitive to the lattice degree of freedom may be needed to fully understand the role of the lattice component in the CDW dynamics.

We thus recognize the melting process as the removal of the electronic order contribution to the CDWs, uncovering persistent short-range CDW correlations, which we interpret as coexisting to the electronic order. In our simplified two-peak model, we considered the two CDW contributions to remain fixed in correlation length. Within this scenario topological defects of long-range CDW would remain largely unperturbed by direct photo-excitation at these low fluence levels, in agreement with similar results in LESCO \cite{bluschke2024orbital}, thus allowing rapid, sub-ps scale, re-establishment of spatial order.  While this model captures the coarse features of our experimental data, we cannot exclude some finer changes in correlation lengths by each CDWs or possible interactions between long- to short-range CDW correlations. This would require further studies with a higher signal-to-noise ratio (via higher repetition rate), and potentially measuring 2D scattering maps or tr-RIXS spectra for further decomposition.

In summary, our time-resolved X-ray scattering measurements reveal fundamental insights into charge-ordering phenomena in YBCO. By systematically varying the pump fluence, we identify a critical fluence $\Phi$$_\mathrm{C}$ where the system exhibits an abrupt, non-thermal transition where the melting of the electronic order uncover persistent and coexisting short-range CDW correlations. Because of the completely different response to photo-excitation, we intreprept the data via two distinct CDW components, short- and long-ranged. 
Our findings corroborate previous RIXS observations of coexisting short- and long-range charge correlations in YBCO \cite{arpaia2019,arpaia2023,arpaia2021charge,chaix2017dispersive,Silva2018coupling,yu2020unusual,miao2017precursor,miao2019formation,li2020multiorbital,lee2021spectroscopic,boschini2021dynamic,huang2021quantum,lu2022identification,scott2023low} and are broadly consistent with short-range charge correlations identified via phonon-tracking measurements in Bi-based cuprates \cite{scott2023low}. While those studies required careful lineshape decomposition to separate the two components at equilibrium, our time-resolved approach provides a new way to discriminate these charge correlations, by exploiting their dramatically different sensitivity to photo-excitation. The results suggest that the two CDW components are not merely different manifestations of a single charge instability, but arise from distinct mechanisms with fundamentally different timescales and stability against photo-excitation. Our approach also allows to clearly recognize the energy threshold for the collapse of long-range CDW order, which is dominated by electronic processes \cite{castellani1995singular}.

\begin{acknowledgements}
We thank Alfred Zong, Matteo Mitrano, Jure Demsar and Daniel Jost for useful discussions. Use of the Linac Coherent Light Source (LCLS), SLAC National Accelerator Laboratory, is supported by the U.S. Department of Energy, Office of Science, Office of Basic Energy Sciences under Contract No. DE-AC02-76SF00515. The SXR Instrument is funded by a consortium whose membership includes the LCLS, Stanford University through the Stanford Institute for Materials Energy Sciences (SIMES), Lawrence Berkeley National Laboratory (LBNL), University of Hamburg through the BMBF priority program FSP 301, and the Center for Free Electron Laser Science (CFEL). This research was undertaken thanks in part to funding from the Max Planck–UBC–UTokyo Centre for Quantum Materials and the Canada First Research Excellence Fund, Quantum Materials and Future Technologies. This project is also funded by the Natural Sciences and Engineering Research Council of Canada (NSERC), the Canada Foundation for Innovation (CFI); the British Columbia Knowledge Development Fund (BCKDF); the Department of National Defence (DND); the Canada Research Chairs Program (F.B., A.D.), the Fonds de recherche du Qu´ebec - Nature et Technologies (F.B.), the Alfred P. Sloan Foundation (F.B.), and the CIFAR Quantum Materials Program (A.D.). E.H.d.S.N. acknowledges prior support from the Alfred P. Sloan Fellowship in Physics and the National Science Foundation under grant nos. 1845994 and 2034345. G.C., C.S. and data analysis work were supported by the Department of Energy, Laboratory Directed Research and Development program at SLAC National Accelerator Laboratory, under contract DE-AC02-76SF00515.
\end{acknowledgements}

\bibliographystyle{apsrev4-2}
\bibliography{Reference}

\end{document}